\documentclass[12pt,preprint]{aastex}
\begin{document}
\title{Testing the Curvature Effect and Internal Origin of
Gamma-Ray Burst Prompt Emissions and X-ray Flares with Swift Data}
\author{E. W. Liang$^{1,2}$, B. Zhang$^1$, P. T. O'Brien$^3$,
R. Willingale$^3$, L. Angelini$^4$, D. N. Burrows$^5$, S. Campana$^6$,
G. Chincarini$^{6,7}$, A. Falcone$^5$,  N. Gehrels$^4$,
M. R. Goad$^3$, D. Grupe$^5$, S. Kobayashi$^8$,
P. M\'esz\'aros$^{5,9}$, J. A. Nousek$^5$,
J. P. Osborne$^3$, K. L. Page$^3$, G. Tagliaferri$^6$}
\affil{$^1$Physics Department, University of Nevada,
Las Vegas,
NV 89154; \\lew@physics.unlv.edu, bzhang@physics.unlv.edu.\\
$^2$Department of Physics, Guangxi University, Nanning 530004, China. \\
$^3$Department of Physics and Astronomy, University of Leicester,
Leicester LE1 7RH, UK.\\
$^4$NASA Goddard Space Flight Center, Greenbelt, MD 20771.\\
$^5$Department of Astronomy and Astrophysics, Pennsylvania State
University, University Park, PA 16802. \\
$^6$INAF-Osservatorio Astronomico di Brera, Via Bianchi 46, I-23807
Merate, Italy. \\
$^7$Universit\'a degli studi di Milano-Bicocca, Dipartimento di
Fisica, Piazza delle Scienze 3, I-20126 Milan, Italy. \\
$^8$Astrophysics Research Institute, Liverpool John Moores University,
Twelve Quays House, Birkenhead, CH41 1LD, UK.\\
$^9$ Department of Physics, Pennsylvania State University, University
Park, PA 16802
 }

\begin{abstract}
The X-ray light curves of many gamma-ray bursts (GRBs) observed by the
{\em Swift} X-Ray Telescope (XRT) have a very steep-decay component
(tail) following the prompt gamma-rays in the early phase and have
some erratic flares occurring at a time from $\sim 10^2$ up to $\sim
10^5$ seconds. Based on the assumption that these prompt emission
tails and flares are
of ``internal'' origin and that their decline behaviors are
dominated by the curvature effect of the fireball, we present a
self-consistency test for this scenario with a sample of 36
prompt-emission-tails/flares in 22 GRB XRT light curves. The
curvature effect suggests that the temporal decay slope of the late
steep-decay part of the light curves is $\alpha=2+\beta$, where
$\beta$ is the X-ray spectral index. We derive the zero time ($t_0$)
for each steep decay component by fitting the light curves with the
constraint of $\alpha=2+\beta$. Our results show that the $t_0$'s of
the prompt emission tails and the tails of well-separated flares are
usually at the rising
segment of the last pulse of the prompt emission or the corresponding
X-ray flare, being self-consistent with the expectation of the
internal dissipation models for the prompt emission and X-ray
flares. Our results indicate that each X-ray flare forms a distinct
new episode of central engine activity and the GRB central engine
remains active after the prompt emission is over, sometimes up to
$\sim$ 1 day after the GRB trigger (e.g. GRB 050502B \& GRB
050724). This challenges the conventional central engine models and
calls for new ideas to re-start the central engine. We further show
that the on-set time of the late central engine activity does not
depend on the GRB duration. We also identify a minority group of GRBs
whose combined BAT-XRT light curves are smoothly connected without an
abrupt transition between the prompt emission and the afterglow. These
GRBs may have an external origin for both the prompt emission and the
afterglow.
\end{abstract}


%%%%%%%%%%%%%%%%%%%%%%%%%%%%%%%%%%%%%%%%%%%%%%%%%%%%%%%%%%%%%%%%
\section {Introduction}
The successful launch and operation of the {\em Swift} mission
(Gehrels et al. 2004) have led to several important discoveries
(e.g. Tagliaferri et al. 2005; Burrows et al. 2005; Gehrels et
al. 2005; Fox et al. 2005;
Barthelmy et al. 2005; Roming et al. 2006; Cusumano et
al. 2006a). Combined analyses of the early data from the Burst Alert
Telescope (BAT) and the X-Ray Telescope (XRT) for a large sample of
bursts (Nousek et al. 2006; O'Brien et al. 2006) reveal a canonical
X-ray afterglow lightcurve characterized by 5 components (Zhang et
al. 2006a, see also Nousek et al. 2006): a steep-decay component
associated with the GRB prompt emission tail, a shallow decay
component likely due to refreshing of the forward shock, a normal
decay component, a possible steep decay component following a jet
break, as well as one or more X-ray flares.  These new data provide
unprecedented information to unveil the nature of these mysterious
explosions.

One of the outstanding problems in the pre-{\em Swift} era concerned
the emission site of the prompt $\gamma$-ray emission. It is generally
believed that GRB prompt emission originates at a distance internal to
the fireball deceleration radius. The most widely discussed scenario
is the internal shock model (Rees \& M\'esz\'aros 1994; Kobayashi et
al. 1997; Daigne \& Mochkovitch 1998), but magnetic dissipation at an
internal radius is also possible (e.g. Usov 1992; Thompson 1994;
Drenkhahn \& Spruit 2002; Giannios \& Spruit 2006). The broad-band
afterglows, on the other hand, are produced by the external shocks
when the fireball is decelerated by the ambient medium (M\'esz\'aros
\& Rees 1997; Sari et al. 1998). This ``internal+external shock" model
suggests that the prompt emission and the afterglow involve two
distinct processes at two different emission sites. Alternatively, it
has been argued that both the GRB prompt emission and their afterglows
are produced in external shocks, provided that the immediate medium
near the burster is clumpy enough (Dermer \& Mitman 1999; 2003). The
evidence collected in the pre-{\em Swift} era cannot conclusively
differentiate between the internal and the external models (see Zhang
\& M\'es\'aros 2004 for a critical review on the successes and
limitations of both models). It is one of the scientific goals of the
{\em Swift} to pin down the emission site of GRB prompt emission using
early afterglow data.

The steep-decay component commonly existing in early X-ray afterglows
(Tagliaferri et al.  2005; Barthelmy et al. 2005b; Cusumano et
al. 2006b; Vaughan et al. 2006) has been generally interpreted as the
tail of the prompt emission (Zhang et al. 2006a; Nousek et al. 2006;
Panaitescu et al. 2006; Yamazaki et al. 2006; Lazzati \& Begelman
2006). This component strongly suggests that the prompt emission and
the afterglow are two distinct components, which supports the internal
origin of the prompt emission (Zhang et al.  2006a; cf. C. D. Dermer,
2006, in preparation). The distinct X-ray flares typically show rapid
rise and fall, with the ratio of the variability time scale and the
epoch of the flare typically much less than unity, i.e. $\delta t/t
\ll 1$ (Burrows et al. 2005; Falcone et al. 2006; Romano et
al. 2006). Burrows et al. (2005) proposed in their discovery paper
that the flares are also produced in internal shocks at later times,
which requires reactivation of the GRB central engine. Zhang et
al. (2006a) performed more detailed analysis of various possible
scenarios and concluded that the late internal dissipation model is
the correct interpretation of X-ray flares. Similar conclusions have
been also drawn by Ioka et al. (2005), Fan \& Wei (2005), Falcone et
al. (2006) and Romano et al. (2006) (c.f. Piro et al.  2005;
C. D. Dermer 2006, in preparation).

An important clue to diagnose the internal origin of the prompt
emission and the X-ray flares is the steep decay components following
the prompt emission and the flares. These mark the sudden cessations
of the emission, and the rapid decays are due to the observer
receiving the progressively delayed emission from higher latitudes -
the so-called ``curvature effect'' (Kumar \& Panaitescu 2000; Dermer
2004; Zhang et al. 2006a; Fan \& Wei 2005; Panaitescu et al. 2006;
Dyks et al. 2006; Wu et al. 2006). The internal emission could be
either from conventional internal shocks (Rees \& M\'esz\'aros 1994;
Kobayashi et al. 1997; Daigne \& Mochkovitch 1998) or from magnetic
dissipation at an internal radius (e.g. Usov 1992; Thompson 1994;
Drenkhahn \& Spruit 2002; Giannios \& Spruit 2006). Usually the
detected emission contains the contributions from many emission
episodes (e.g. emission from many internal shocks or magnetic
dissipation regions). Each emission episode is expected to be followed
by a curvature effect tail after the cessation of the emission
(e.g. shock crosses the shell or magnetic reconnection finishes). For
highly overlapping emission episodes (e.g. the prompt emission), a
curvature tail is usually buried beneath the rising lightcurve of
another episode, so that it is difficult to observe a clear curvature
effect signature. Although extensive studies have been made of the
curvature effect in the prompt emission phase (Fenimore et al. 1996;
Sari \& Piran 1997; Ryde \& Petrosian 2002; Norris 2002; Kocevski et
al.  2003; Qin et al. 2004; Shen et al. 2005; Qin \& Lu 2005), no
conclusive evidence supporting this effect has been presented
(e.g. Kocevski et al. 2003).

A clear, testable prediction of the curvature effect when the viewing
angle is larger than the $1/\Gamma$ cone (where $\Gamma$ is the bulk
Lorentz factor of the emission region) is that the temporal decay
index $\alpha$ should be related to the spectral index $\beta$ by the
expression (Kumar \& Panaitescu 2000)
\begin{equation}
\alpha=2+\beta,
\label{curvature}
\end{equation}
where the convention $F_\nu \propto t^{-\alpha} \nu^{-\beta}$ is
adopted. This relation is valid under the following assumptions,
i.e. (1) there is a sharp drop off in the injection of accelerating
electrons at a certain radius, after which the electrons cool
adiabatically; (2) no spectral break energy crosses the observational
band during the epoch of decay; and (3) the observational band is
above the cooling frequency. When these conditions are satisfied, such
a relation is rather robust regardless of the fireball history
(e.g. Zhang et al. 2006a; Fan \& Wei 2005; Dyks et al. 2006; Wu et
al. 2006), and depends only on the jet structure when the line of
sight is outside the bright jet core (Dyks et al. 2006).

Several effects could lead to deviations from eq.(1). The most
important one is the zero time effect ($t_0$-effect). The $t_0$-effect
is crucial, since when considering the
multiple reactivation of the central engine, the GRB trigger time is
no longer special (Zhang et al. 2006a). Every time when the central
engine is re-started, the new central engine time should be re-set as
$t_0$. The conventional lightcurves are plotted as $\log F_\nu$ -
$\log(t-t_{\rm trigger})$, where $t_{\rm trigger}$ is the time when
the GRB triggers BAT. When we consider the emission from a late
central engine activity episode, the relevant decay slope (in order to
be compared with the theoretical prediction in eq.[1]) is $d \ln
F_\nu/ d \ln (t-t_0)$ rather than $d \ln F_\nu/ d \ln (t-t_{\rm
trigger})$. Properly shifting $t_0$ is therefore crucial to understand
the real temporal decay index in the lightcurves\footnote{The $t_0$
issue is also relevant when discussing orphan afterglows from dirty
fireballs (Huang et al. 2002).}. For internal
dissipation models (both internal shocks or internal magnetic
dissipation), the expected $t_0$ of a certain flare should be at the
beginning of the corresponding emission episode, i.e. near the
starting point of the rising segment of the flare. Since both $\alpha$
and $\beta$ could be directly derived from the Swift XRT observations,
this provides a solid self-consistency test to the curvature-effect
interpretation and the internal-origin hypothesis.

Another effect that leads to deviations from eq.(1) is the overlapping
effect (Zhang et al. 2006a), i.e. there is an underlying forward shock
component beneath the steep decays.  In order to perform a clean test
to the curvature effect, such a component needs to be subtracted.
Kumar \& Panaitescu (2000) suggested that the contribution of the
external shock component is significant if the ambient density is not
lower than $10^{-2}~{\rm cm}^{-3}$.

In this paper we use the XRT data to test the curvature-effect
interpretation and internal-origin hypothesis of the GRB prompt
emission and X-ray flares. We take the GRB sample presented by O'Brien
et al. (2006) and focus on the steep decay components following the
prompt emission and the X-ray flares. We {\it assume} that the
curvature effect is the cause of the steep decays so that
eq.(\ref{curvature}) is valid ubiquitously. After subtracting the
underlying contribution from the forward shock, we derive the $t_0$
values by fitting the light curves with our model. The object is to
check whether the location of $t_0$ is consistent with the expectation
of the model, i.e.  near the starting point of the relevant emission
episode. The fitting model is described in \S 2. The data and the
fitting results are presented in \S 3. Conclusions are drawn in
\S4 with some discussion.

\section{Model}
The rapid decay component is our primary interest. As discussed above
we assume that this component is mainly contributed by the curvature
effect. On the other hand, the overlapped contribution from an
external shock component for $n>10^{-2}$ cm$^{-3}$ could result in a
decay that would deviate significantly from eq.(1) (Kumar \&
Panaitescu 2000). Considering both the $t_0$-effect
and the overlapping effect, we model any steep decay component and the
succeeding afterglow component with the function (Zhang et al. 2005)
\begin{equation}
F_{\nu}(t)=A\left(\frac{t-t_0}{t_0}\right)^{-(2+\beta)}+Bt^{-C}~,
\label{fits}
\end{equation}
where $\beta$ is the X-ray spectral index during the decay, $t_0$ is
the time zero point of the emission episode related to the decay
(which in principle should be at the beginning of the rising segment
of the last pulse of the prompt emission or the relevant X-ray
flare if the curvature interpretation is correct), $A$ and $B$ are
normalization parameters for both the rapid
decay component and the underlying forward shock component,
respectively, and $C$ is the temporal index of the forward shock
emission component. Please note that we take the zero point time of
the external shock component (afterglow) as the GRB trigger time. This
has been proved by detailed numerical simulations (Lazzati \& Begelman
2006; S. Kobayashi et al. 2006, in preparation).

\section{Data and fitting results}
The {\em Swift} GRB sample we use is the same sample presented by
O'Brien et al. (2006).  These are the bursts detected by the {\em
Swift} prior to October 1, 2005 for which prompt slews within 10
minutes were performed. The BAT data were processed using the standard
BAT analysis software ({\em Swift} software v. 2.0). The XRT data were
processed using {\tt xrtpipeline v0.8.8}. The joined BAT-XRT
lightcurves were derived through extrapolating the BAT lightcurves
into the XRT band (0.3-10 keV) using the joint BAT-XRT spectral
parameters. For the details of the data reduction procedure we refer
to O'Brien et al. (2006). For our purpose, we include only those GRBs
whose XRT lightcurves have a steep decay component connecting to the
prompt emission and those GRBs that harbor X-ray flares. In some
bursts there are multiple flares, and hence, multiple steep
decays. For these bursts we treat each steep decay independently. We
consider only those prompt emission tails or flare tails whose decay
slopes are steeper than -2 with the zero time set to the GRB trigger
time. For the heavily-overlapped flares in the early XRT light curves,
we take only those that are well identified visually and without
significant substructures. Our sample includes 36
prompt-emission/flare tails from 22 GRBs, which are tabulated in Table
1. Their lightcurves are collected in Fig.1.

Technically one needs to identify the time interval for which our
above test is performed. Since the steep-decay components (the tails
of the prompt emission or the X-ray flares) are required to satisfy
the curvature effect, we need to search for a segment of lightcurve
whose decay slope is steeper than -2. We choose a time window that
contains 4 data points. We start the search from the beginning of the
steep decay that follows the prompt emission or from the peak of a
certain flare. We then move the time window to later times by shifting
one data point in each step. We fit the decay slope for the four data
points in the time window in each step until the slope becomes steeper
than -2. By then the beginning of the time window is set to the
starting point of the steep decay component. The end of the time
interval for which our fit is performed is selected visually without a
rigid criterion.  For those prompt-emission/flare tails with an
underlying shallow decay (afterglow) component, we take the end of
that component (before a further break if any) as the end of the time
interval and fit the data using eq.(\ref{fits}). For some tails
without superposition of an underlying afterglow component or
otherwise highly overlapped with other flares, we simply choose the
last data point of the steep decay component as the end of the time
interval and fit the lightcurve using eq.(\ref{fits}) by setting
$B=0$.

The GRBs 050406, 050502B, 050713B, and 050801 have a very flat segment
following the tails of the prompt emission or flares. These flat
segments have only a few data points, and we fix the $C$ values in our
fitting for these GRBs. Several bursts, e.g. GRBs 050712 (the first
and the second flares), 050713A, 050716 (the first flare), 050730 (the
1st - 4th flares), and 050822 (the first and second flares) have heavily
overlapped flares before the typical long tail as observed in many
other GRBs (e.g. 050126, 050219, 050315, etc.).  We also fit these
flares. The XRT light curve of GRB 050714B has a rapid flare in the
tail segment. The level of this flare is significantly higher than the
tail. We subtract the flare from the tail segment and fit the tail
segment and the flare independently.

The spectral indices we use are from O'Brien et al. (2006). These
spectral indices are derived from the spectral fitting to the overall
X-ray data without considering spectral variability. In principle one
should use the $\beta$ value during the steep decay to perform the
test. However the photon counts during the steep decay only are
usually too low to give a high significance fit. The X-ray spectra of
some GRBs, such as GRB 050502B (Falcone et al. 2006) and GRB 050607
(Pagani et al 2006), have detectable spectral variability. We check
the difference made by this effect with GRB 050502B, and find that
this effect does not significantly affect our fitting
results. Throughout this analysis we use the spectral indices from
O'Brien et al. (2006).

Our fitting results are summarized in Table 1.  In Figure 1, we mark
the time interval of the fitting data and the fitting curve by solid
lines, and indicate the fitted $t_0$ of each steep decay component by
a vertical dash-dotted line. For those lightcurves with multiple
tail/flares, we identify the $t_0$ of the $i$-th steep decay component
as $t_{0,i}$. The reduced $\chi^2$ of the fits are also shown in
Fig. 1.

From Fig. 1 we find that, except those heavily-overlapped flares in
the early XRT light curves, the fitted $t_0$'s of a good fraction of
the well-identified tails are right at the beginning of the
rising phase of the flare or the last pulse of the prompt
emission. These include GRBs 050126, 050219A, 050319, 050406, 050422,
050502B (the late flare), 050607, 050712 (the third flare), 050716
(the second flare), 050724 (the late flare), 050803, and 050822 (the
third flare). This is well consistent with our starting hypothesis,
i.e. the flares are of internal origin and mark the re-activations of
the central engine (Zhang et al. 2006a). We emphasize that the fitted
$t_0$'s are based on the hypothesis of the curvature effect as the
origin of the steep decay, which is only relevant to internal models,
not to external models. So even if some external models may also allow
re-set of $t_0$ before the flares, the decay slope after the peak
would follow some other predictions (typically flatter than
$-(2+\beta)$, e.g. Wu et al.  2006). As a result, the fitted $t_0$'s
to satisfy those predictions would be significantly different from the
ones we obtained, i.e. they should have a large off-set with respect
to the rising part of the flare. This means that one does not find a
self-consistent solution for the external shock models. The impressive
consistency displayed in the above bursts lends strong support to the
internal origin of these X-ray flares. It is worth noticing that a
very late flare (around 1 day) is evident in the XRT lightcurves of
GRB 050502B (Falcone et al. 2006) and the short burst GRB 050724
(Barthelmy et al. 2005a). There have been questions as to whether the
central engine can restart at such a late epoch and whether these
features are due to a refreshed external shock origin (e.g. Panaitescu
et al. 1998; Zhang \& M\'esz\'aros 2002). Our results indicate
remarkably that the $t_0$'s for these two late flares are also right
before the rising part of the flares. This gives strong support to the
internal origin of these flares, and calls for central engine models
that can operate as long as $\sim$ 1 day.

The fitted $t_0$'s for some tails in GRBs 050315, 050712, 050713A,
050714B, 050716, 050721, 050724, 050730, 050801, 050814, 050819 and
050822 are all before the peak of the corresponding flare or the
starting point of the steep decay. However, they are not apparently
located at the rising part of a flare or prompt emission pulse.  These
are all highly overlapping flares or continuously decreasing prompt
emission, so that the rising segment of the flare or the prompt
emission pulse is deeply buried beneath the continuous emission. These
cases do not contradict the internal models and the curvature effect
interpretations, although it does not directly support the scenario.
Further complications may come from the possibility that the electron
injection does not cease abruptly, that the observational band is
below the cooling frequency, and that spectral breaks may cross the
band during the decay, etc. In any case, it confirms that it is
difficult to search for evidence of the
curvature effect using the prompt GRB light curves. On the other hand,
if one believes the internal origin and curvature effect scenario (as
is self-consistently tested in other bursts), then the fitted $t_0$'s
in these bursts give some indication of the central engine time when
the flare or the last pulse of the prompt emission is powered.

In Figure 2, we display the fitted $t_0$'s as a function of $T_{90}$,
the duration of the prompt emission. The cases of heavily overlapped
flares are not included. It is found that the two quantities are not
correlated, indicating that the on-set time of the late central engine
activity does not depend on the duration of the prompt central engine
activity. This suggests irregular, unpredictable behaviors of the GRB
central engine.

We have used the lightcurve in the observer's frame to perform the
above fits. Using the intrinsic lightcurves (i.e. systematically
changing the time axis to the cosmic proper frame time $t/(1+z)$)
shifts the searched time zero point to $t_0/(1+z)$. So our conclusion
does not depend on the time dilation effect since the lightcurves are
simply compressed in the cosmic proper frame.

\section{Conclusions and Discussion}
We have analyzed 36 prompt-emission/flare tails in the XRT light
curves of 22 GRBs detected by the {\em Swift} before October 1, 2005
that show clear steep decay components. This is a sub-sample of that
presented by O'Brien et al. (2006). Assuming that the tails are
predominantly caused by the curvature effect, we derive the $t_0$'s of
these tails by fitting the XRT light curves with eq.(2). Our results
(Fig.1 and Table 1) suggest that usually the $t_0$'s are near the
beginning of the rising segment of the last pulse of the prompt
emission or the corresponding X-ray flare, which is consistent with
the expectation of the internal dissipation models for the prompt
emission and X-ray flares (e.g. Burrows et al. 2005; Zhang et
al. 2006a; Fan \& Wei 2005; Ioka et al. 2005; Falcone et al. 2006;
Romano et al. 2006). This suggests
that the GRB central engine re-activates after the early prompt
emission is over, sometimes up to days after the trigger (e.g. GRB
050502B \& GRB 050724). This challenges the conventional central
engine models and calls for new ideas to re-start the central engine
(e.g. King et al. 2005; Perna et al. 2006; Fan et al. 2005; Proga \&
Zhang 2006; Dai et al. 2006). We also show that the on-set times of
the late central engine activity do not depend on $T_{90}$, suggesting
an erratic, unpredictable central engine.

About one-third of the light curves in the O'Brien et al. (2006)
sample do not show a distinct steep-decay component that connects the
prompt emission and the afterglow. It is possible that in some GRBs
(e.g. GRB 050525A, Blustin et al.  2006) the XRT observations started
too late to catch the prompt emission tail.  Nonetheless, a small
fraction of XRT light curves show an apparently smooth transition from
the prompt emission to the afterglow emission without a steep-decay
bridge. The most prominent cases are GRB 050401 (De Pasquale et
al. 2006), 050717 (Krimm et al. 2006), and 050826 (O'Brien et
al. 2006).  The joined BAT-XRT light curves of these GRBs (see O'Brien
et al. 2006) have some low-amplitude flares overlapping on an
otherwise smooth single power-law decay component extending to a late
epoch. If these lightcurves are still interpreted within the frame
work of internal-external model, as is favored by the majority of
bursts in the sample (Fig.1), the prompt emission and the flares are
of internal origin with a smaller emission efficiency than most other
bursts (Zhang et al. 2006b). For example, the lightcurve of GRB 050717
could be also well modelled by the superposition of a prompt emission
tail emission and an underlying afterglow component (Krimm et
al. 2006).  Alternatively, it may be possible that the fireball in
these cases is decelerated at a very early time so that both the
prompt emission and the afterglow originate from the external shocks
(e.g. Dermer \& Mitman 1999; 2003). This requires a high ambient
density and a high initial Lorentz factor of the fireball. The low
amplitude flares overlapping on the decaying lightcurves could be
still from late internal dissipation of the central engine, or from
some external shock related collisions (Wu et al. 2006), or else from
possible density clumps in the medium (e.g. Dermer \& Mitman 1999;
2003).

We thank Z. G. Dai, Y. F. Huang, D. A. Kahn, D. Kocevski, J. Norris
and X. F. Wu for useful discussion/comments.
This work is supported by NASA under NNG05GB67G, NNG05GH92G, and
NNG05GH91G (BZ \& EWL), and the National Natural Science Foundation of
China (No. 10463001, EWL).

\newpage

\begin{deluxetable}{lllllllll}
\rotate

\tabletypesize{\scriptsize} \tablewidth{7.1in}

\tablecaption{\label{Gamma-ray}The GRB sample and the fitting results}

\tablecolumns{9}

\tablehead{

\colhead{GRB\tablenotemark{a}}      &

\colhead{Interval(s)\tablenotemark{b}}    &

\colhead{$\beta$\tablenotemark{c} } &

\colhead{$t_{\rm p}$(s)}  &

\colhead{$A$(erg cm$^{-2}$ $s^{-1}$)} &

\colhead{$t_0$(s)} &

\colhead{$B$(erg cm$^{-2}$ $s^{-1}$)}  &

\colhead{$C$}  &

\colhead{$\chi^2/dof$}
}

\startdata
050126&139-29171&1.59(0.38)&5&1.05(20)$\times 10^{-6}$&$0.18(22)$&4.62(4.46)$\times 10^{-9}$&$0.98(0.10)$&0.71\\
050219A&111-30460&1.02(0.20) &97&1.23(1.35)$\times 10^{-8}$&$36(10)$&1.04(0.47)$\times 10^{-9}$&$0.62(0.05)$&1.74\\
050315&134-6230&1.50(0.40) &25&2.76(1.87)$\times 10^{-9}$&$62(8)$&9.22(7.63)$\times 10^{-12}$&$0.03(0.11)$&0.87\\
050319&250-93922&2.02(0.47) &250&4.23(5.39)$\times 10^{-11}$&$167(20)$&1.07(0.28)$\times 10^{-9}$&$0.52(0.03)$&1.09\\
050406&213-881&1.37(0.25)&213&1.64(1.27)$\times 10^{-11}$&$144(12)$&4.85(0.76)$\times 10^{-12}$&0(fixed)&0.06\tablenotemark{d} \\
050421&97-473&0.27(0.37)&115&5.08(1.83)$\times 10^{-9}$&$43(5)$&-& -& 1.54\\
050422&98.5-296896&2.33(0.60)&53&5.31(5.33)$\times 10^{-8}$&$30(5)$&1.11(0.47)$\times 10^{-9}$&$0.86(0.04)$&0.44\tablenotemark{e}\\
050502B(1)\tablenotemark{f}&818-43500&0.81(0.28)&758&2.23(0.54)$\times 10^{-11}$&$680(16)$&1.06(2.19)$\times 10^{-12}$&$0.0(0.21)$&1.36\\
050502B(2)&67300-202000&0.81(0.28)&73300&2.45(5.21)$\times 10^{-11}$&$19603(11312)$&4.99(3.52)$\times 10^{-14}$&0(fixed)& 0.63\\
050607&321-79473&0.77(0.48)&$321$&1.05(0.48)$\times 10^{-10}$&$238(13)$&4.34(2.54)$\times 10^{-9}$&$0.96(0.09)$&0.61\\
050712(1)&227-242&0.90(0.06)&227&3.78(15.06)$\times 10^{-14}$&$220(9)$&-&-&0.11\tablenotemark{g}\\
050712(2)&269-295&0.90(0.06)&270&2.86(1.41)$\times 10^{-11}$&$208(9)$&-&-&0.43\tablenotemark{g}\\
050712(3)&505-379262&0.90(0.06)&485&1.26(1.23)$\times 10^{-12}$&$440(22)$&1.40(0.49)$\times 10^{-8}$&$0.92(0.04)$&1.67\\
050713A(1)&78-98&1.30(0.07)&65&7.30(4.55)$\times 10^{-9}$&$40(4)$&-&-&1.21\\
050713A(2)&114-154&1.30(0.07)&112&$1.08(0.23)\times 10^{-9}$&$79(2)$&-&-&1.5\\
050713A(3)&172-230&1.30(0.07)&171&8.90(4.70)$\times 10^{-10}$&100(8)&- &-&1.59\\
050713B&150-5630&0.70(0.11)&17&9.41(7.87)$\times 10^{-8}$&$31(8)$&1.94(0.44)$\times 10^{-11}$&$\sim 0.01$ (fixed)&1.46\\
050714B(1)\tablenotemark{h}&158-393361&4.50(0.70)& 41&2.89(10.84)$\times 10^{-3}$&$21(11)$&1.57(1.05)$\times 10^{-9}$&$0.57(0.06)$& 1.31\\
050714B(2)\tablenotemark{i}&401-481&4.50(0.70)&400&2.12(5.70)$\times 10^{-7}$&$142(39)$&-&-&0.77\\
050716(1)&177-345&0.33(0.03)&177&2.08(2.85)$\times 10^{-7}$&$22(12)$&-&-&1.34\\
050716(2)&375-244935&0.33(0.03)&383&2.69(1.36)$\times 10^{-10}$&$224(22)$&2.59(1.07)$\times 10^{-8}$&$1.00\pm 0.04$&1.18\\
050721&195-244289&0.74(0.15)&1.5&3.47(11.56)$\times 10^{-8}$&$38(37)$&2.33(0.70)$\times 10^{-7}$&1.18(0.03) &1.36\\
050724(1)&186-7097&0.95(0.07)&73&2.54(1.98)$\times 10^{-8}$&$63(12)$ &1.14(3.90)$\times 10^{-8}$&$1.13(0.40)$&1.70\\
050724(2)&70000-99000&0.95(0.07)&58868&7.25(11.08)$\times 10^{-10}$&$8752(4047)$&-&-&0.02\tablenotemark{g}\\
050730(1)&135-201&0.33(0.08)&140&4(75)$\times 10^{-5}$&$2(16)$&-&-& 1.00\\
050730(2)&230-312&0.33(0.08)&230&3.20(2.84)$\times 10^{-9}$&$95(23)$&-&-& 2.00\\
050730(3)&435-600&0.33(0.08)&430&1.49(0.48)$\times 10^{-9}$&$227(17)$&-&-& 1.80\\
050730(4)&684-793&0.33(0.08)&680&1.79(1.60)$\times 10^{-9}$&$309(69)$&-&-& 1.73\\
050730(5)&10000-120000&0.33(0.08)&4700&7.94(5.50)$\times 10^{-9}$&$1540(407)$&-&-&2.34\\
050801\tablenotemark{j}&17-109&0.72(0.54)&17&8.62(20.93)$\times 10^{-7}$&$2.7(2.1)$&2.68(2.49)$\times 10^{-11}$&0.1(fixed)&0.27\tablenotemark{g}\\
050803&160-1000&0.71(0.16)&145&2.47(1.19)$\times 10^{-10}$&$107(7)$&6.50(8.36)$\times 10^{-10}$&$0.36(0.20)$&1.27\\
050814&166-58886&1.08(0.08)&8&2.53(0.34)$\times 10^{-6}$&$15(6)$&1.18(0.57)$\times 10^{-9}$&$0.72(0.05)$& 1.30\\
050819&33-11369&1.18(0.23)&10&1.56(3.00)$\times 10^{-8}$&$33(17)$&2.02(6.38)$\times 10^{-11}$&$0.44(0.36)$&1.53\\
050822(1)&165-200&1.60(0.06)&130&2.13(1.53)$\times 10^{-9}$&$85(9)$&-&-&1.25\\
050822(2)&236-293&1.60(0.06)&238&3.15(1.47)$\times 10^{-10}$&$145(8)$&-&-&1.06\\
050822(3)&490-16870&1.60(0.06)&450&6.70(1.26)$\times 10^{-11}$&$354(6)$&5.78(4.40)$\times
10^{-11}$&$0.23(0.08)$&0.84
\enddata
\tablenotetext{a}{The number following the burst name denotes the
sequence number of the steep decay components in the light curve
according to the time sequence.}

\tablenotetext{b}{The time interval relative to the GRB trigger time for the data we use
for the fitting, in units of seconds.}

\tablenotetext{c}{Taken from O'Brien et al. 2006. These indices are derived from the
spectral fitting to the overall X-ray data without considering the spectral variability.}

\tablenotetext{d}{The decay part of the flare and the succeeding flat component have only
four data points with large error bars. We fix the $C$ value and fit the data. The error
of $t_0$ is thus small with a small reduced $\chi^2$.}

\tablenotetext{e}{If the last data point at 704682 seconds is considered the reduced
would be then $\chi^2\sim 1$. This point drives most of the $\chi^2$ values.}

\tablenotetext{f}{Some observational data points of this burst have an extremely small
error, say, $\Delta \log F_X<0.1$. The $\chi^2$ of the fitting is mostly contributed by
these data points, and the reduced $\chi^2$ is unacceptable. To make the fitting results
more reasonable we take $\Delta \log F_X=0.1$ for those data points with $\Delta \log
F_X<0.1$. }

\tablenotetext{g}{These flares have few data points with large error
bars. Their reduced $\chi^2$ is too small. The fitting results are
highly uncertain.}

\tablenotetext{h}{Excluding the superimposed X-ray flare at 331-477
seconds.}

\tablenotetext{i}{Fit to the superimposed X-ray flare at 331-477
seconds only.}

\tablenotetext{j}{Fit to the combined BAT-XRT data in 18-110
seconds. The flare at $\sim 210$ seconds does not satisfy our
selection criteria (see in the text).}

\end{deluxetable}

\begin{figure}
\includegraphics[angle=0,scale=.50]{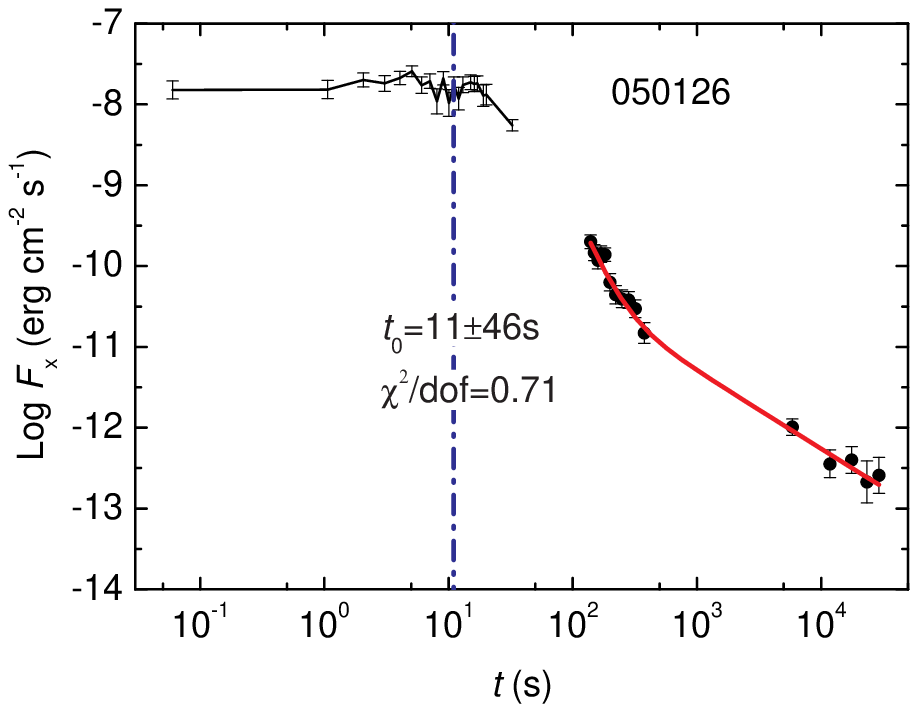}
\includegraphics[angle=0,scale=.50]{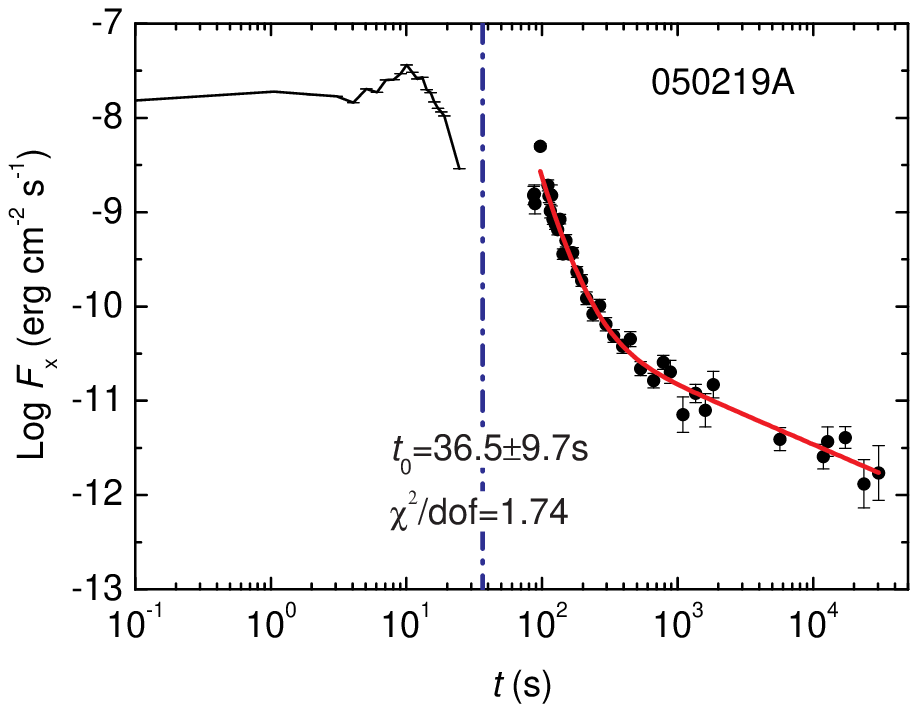}
\includegraphics[angle=0,scale=.50]{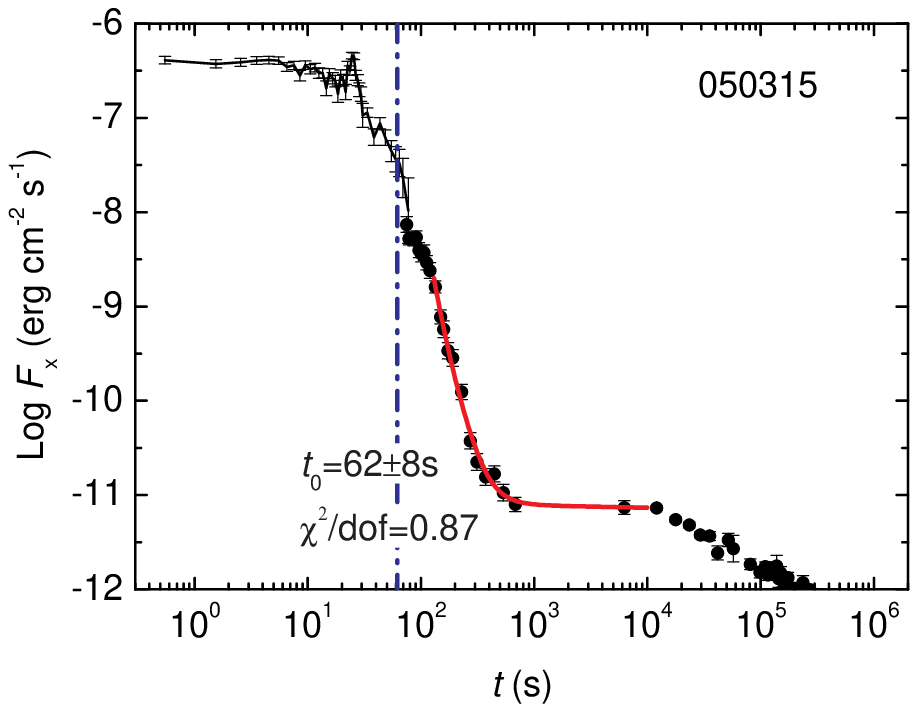}
\includegraphics[angle=0,scale=.50]{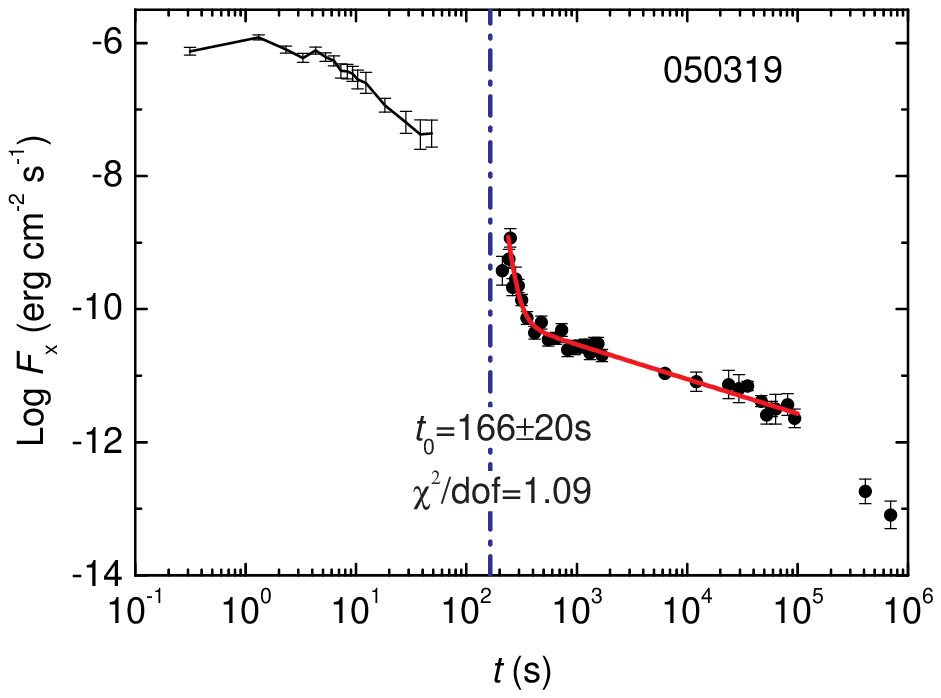}
\includegraphics[angle=0,scale=.50]{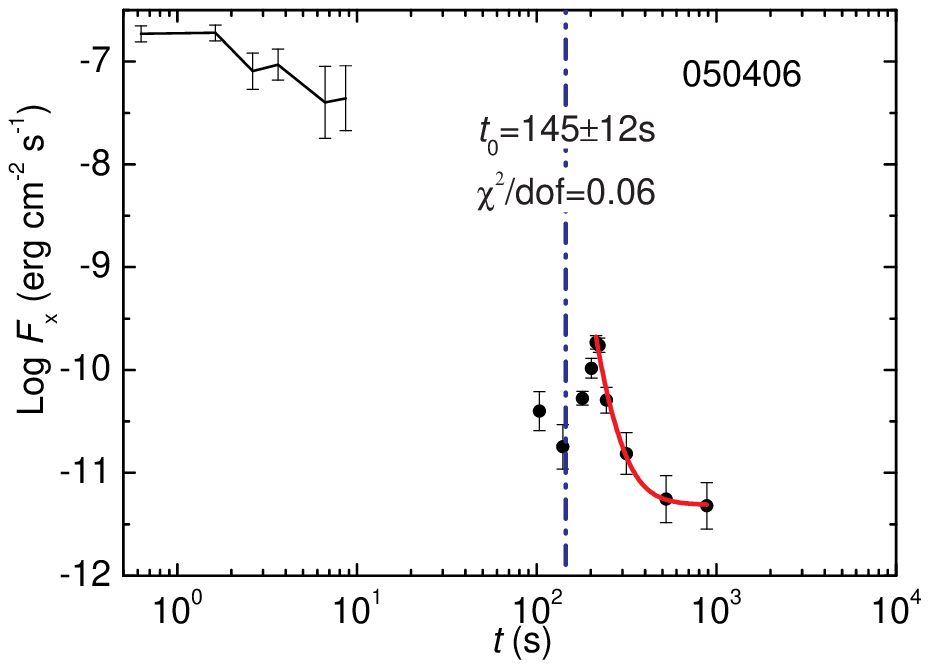}
\includegraphics[angle=0,scale=.50]{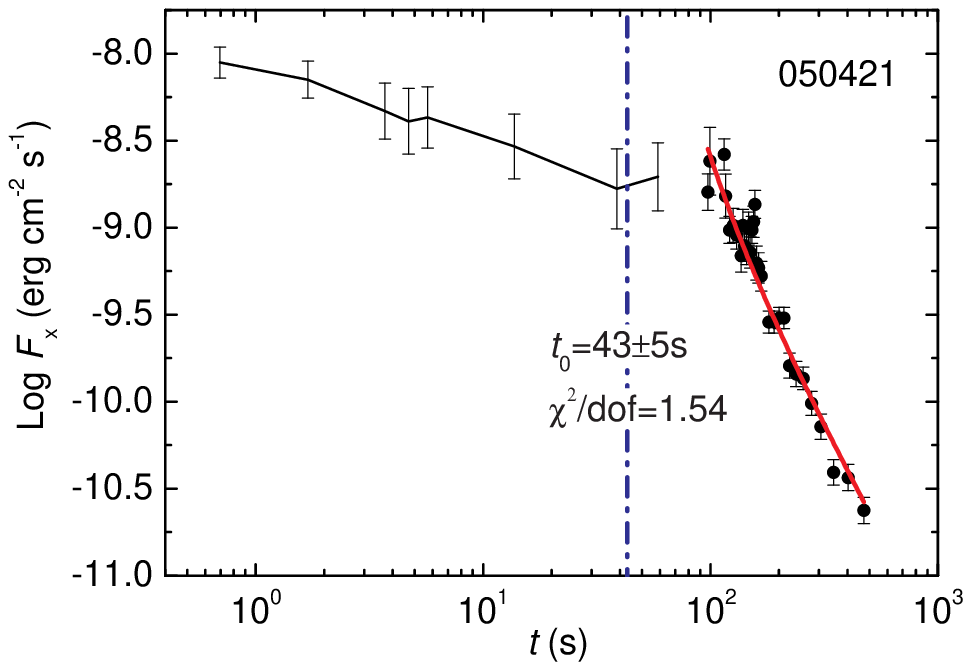}
\includegraphics[angle=0,scale=.50]{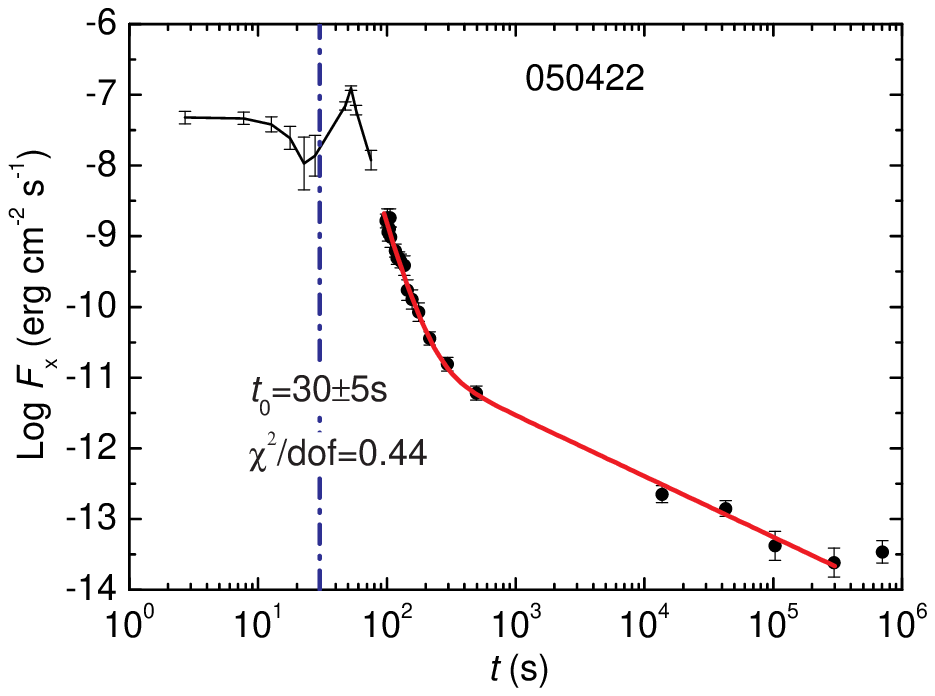}
\includegraphics[angle=0,scale=.50]{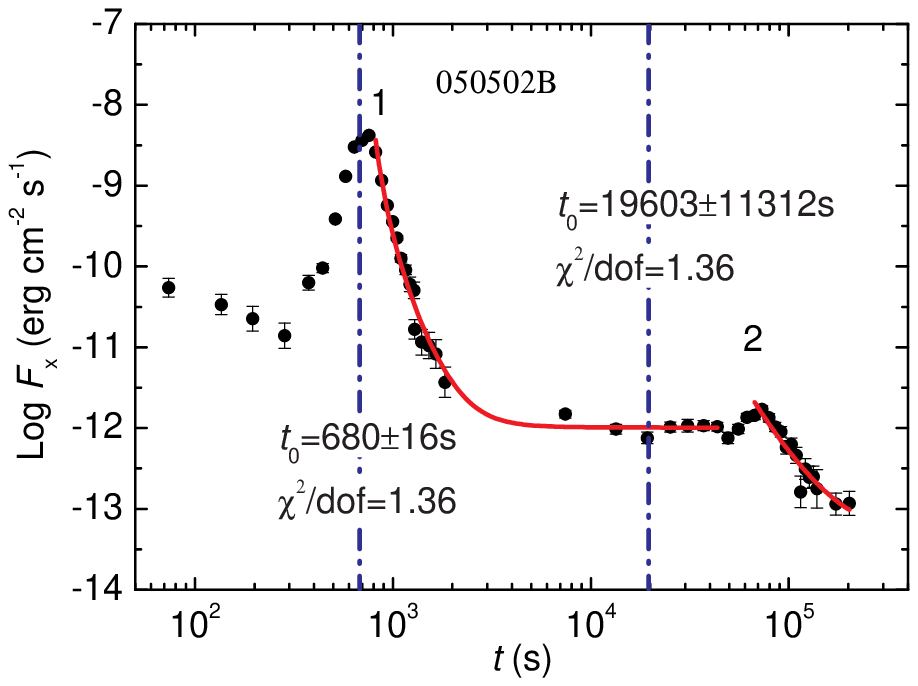}
\includegraphics[angle=0,scale=.50]{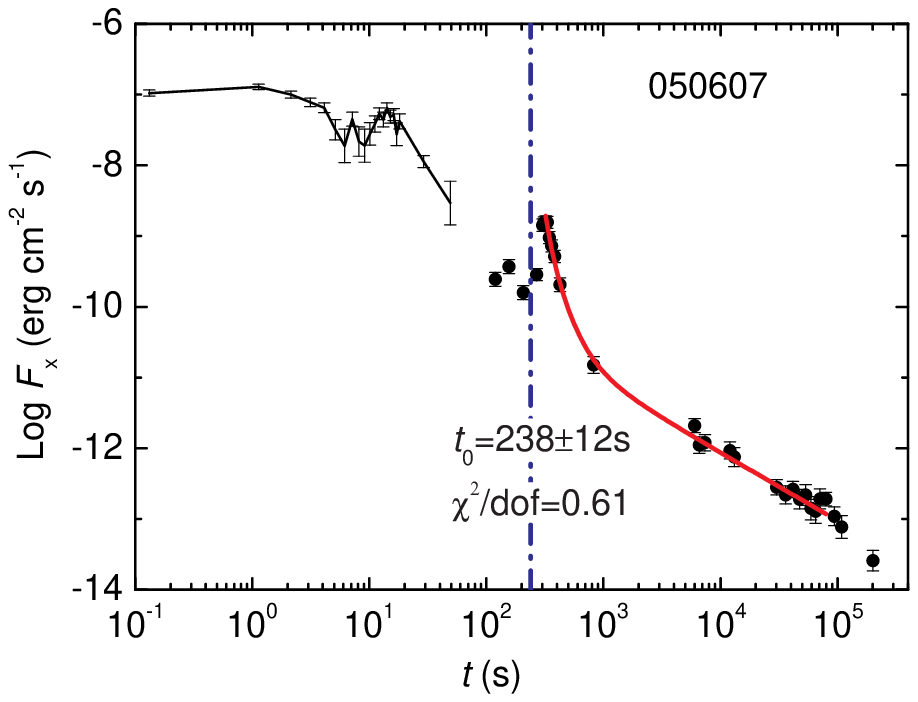}
\includegraphics[angle=0,scale=.47]{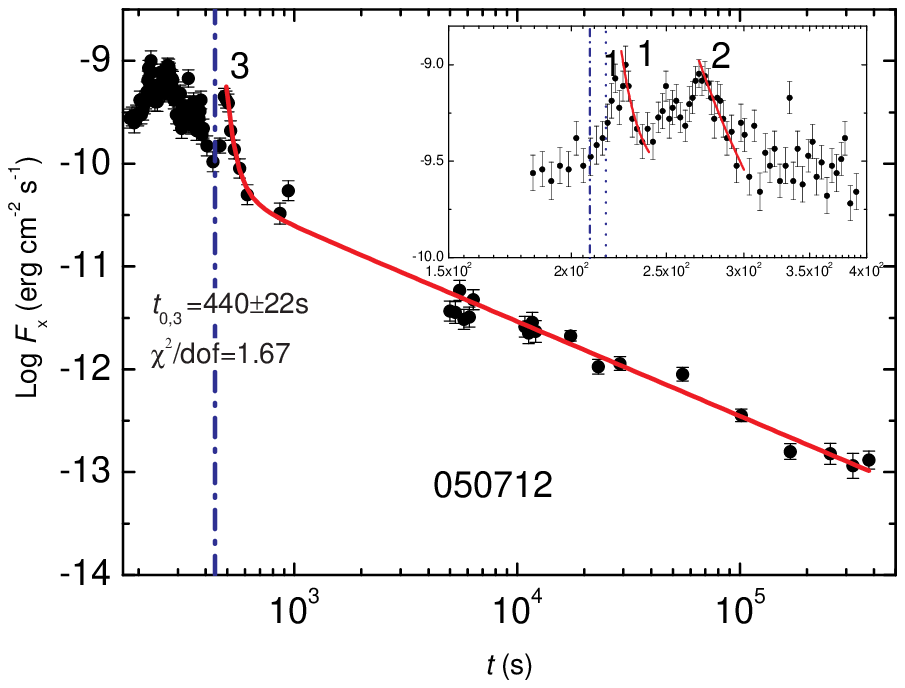}
\includegraphics[angle=0,scale=.50]{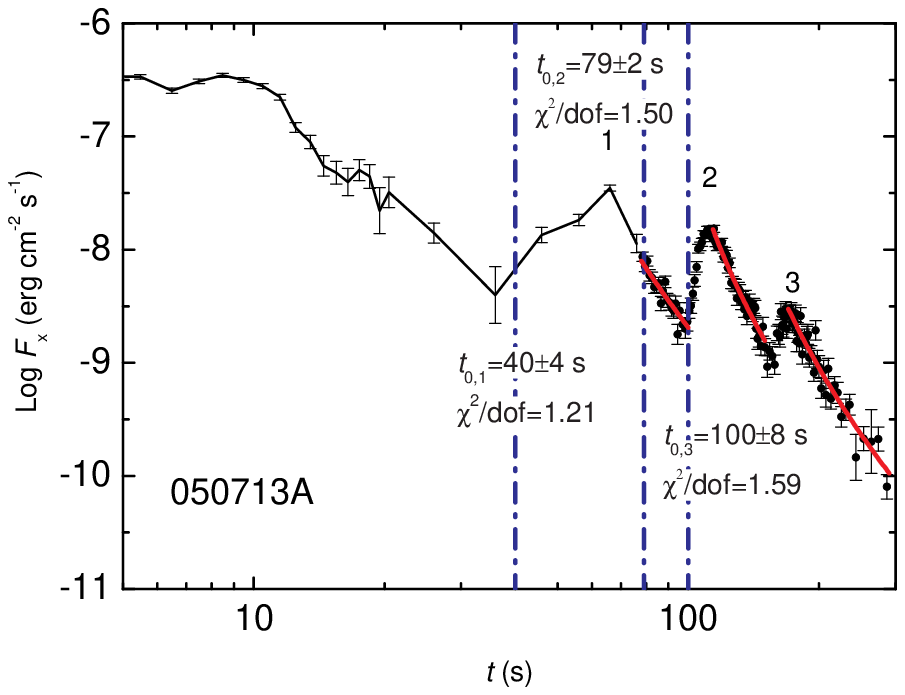}
\includegraphics[angle=0,scale=.50]{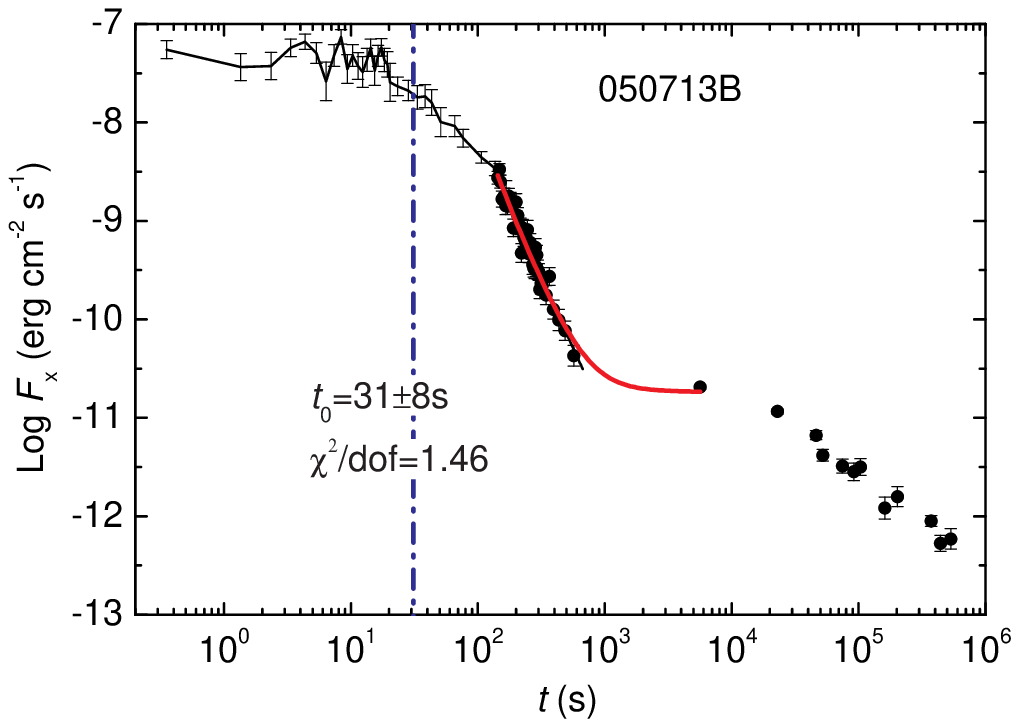}
\caption{The combined BAT-XRT light curves with the best fitting results. The connected
lines with error bars are the extrapolated BAT light curves. The solid circles with error
bars represent the XRT observations. The thick solid curves are the best fits of our
model, which cover the range of the fitted data. The vertical dotted lines mark the best
fit $t_0$'s. When more than one steep decay component is observed, the $t_0$ of the
$i$-th component is denoted as $t_{0,i}$. }
\end{figure}

\begin{figure}

\includegraphics[angle=0,scale=.50]{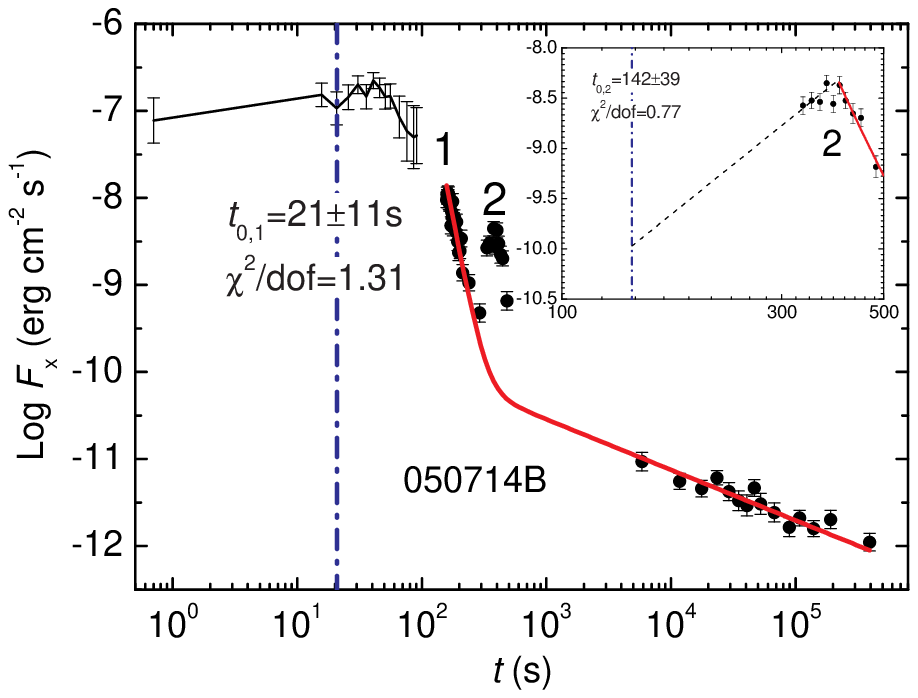}
\includegraphics[angle=0,scale=.47]{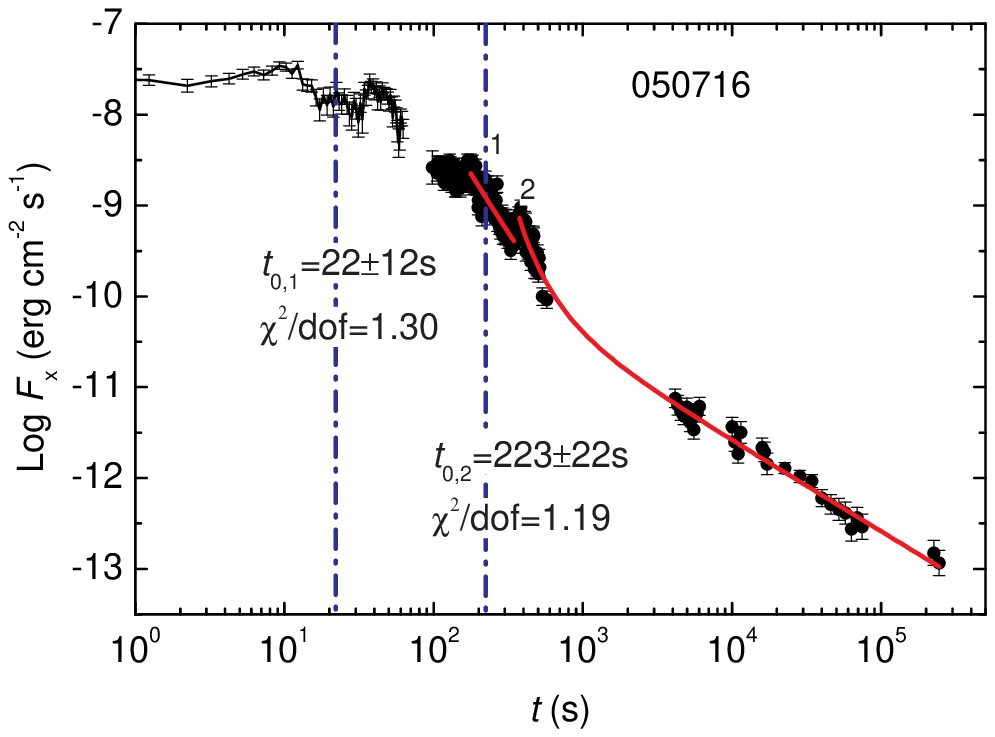}
\includegraphics[angle=0,scale=.50]{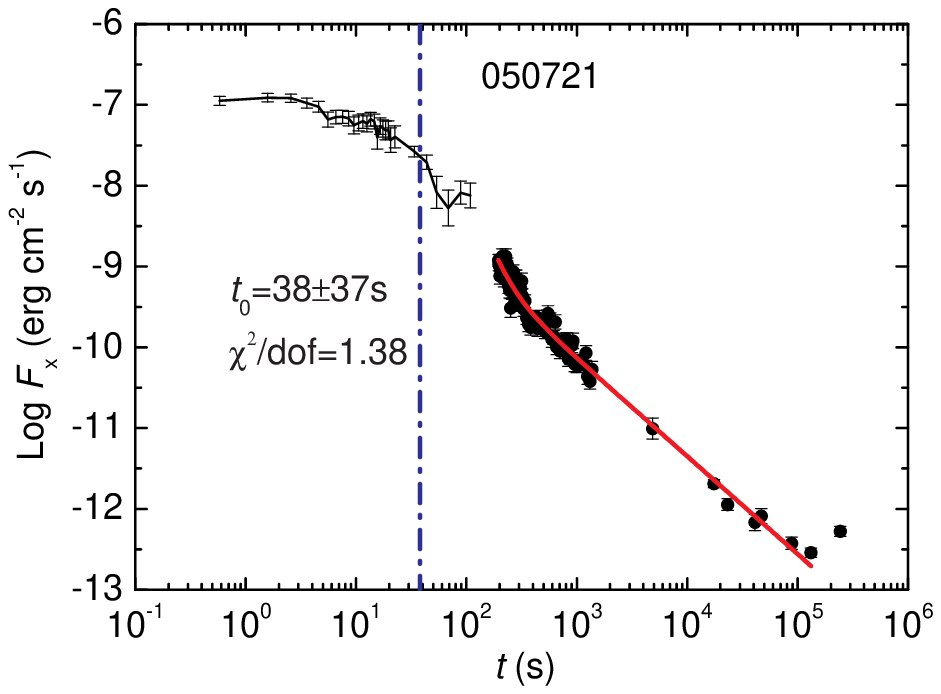}
\includegraphics[angle=0,scale=.50]{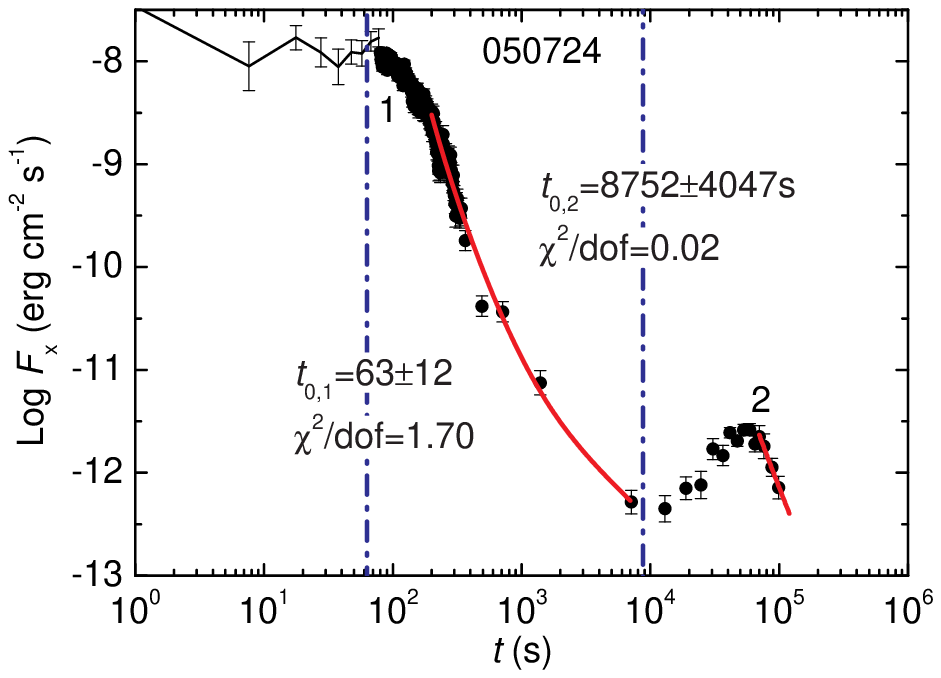}
\includegraphics[angle=0,scale=.50]{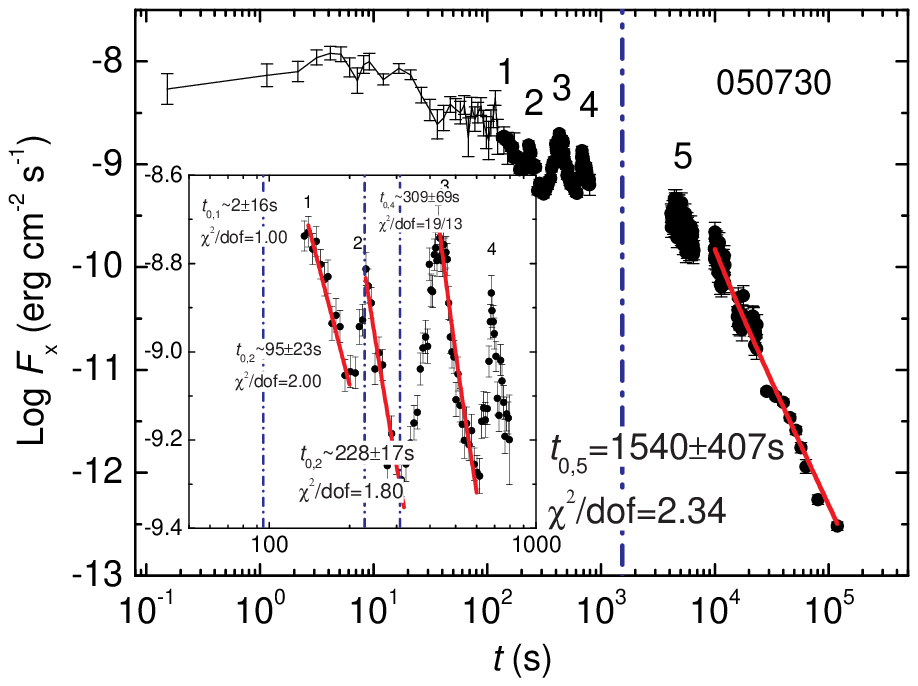}
\includegraphics[angle=0,scale=.50]{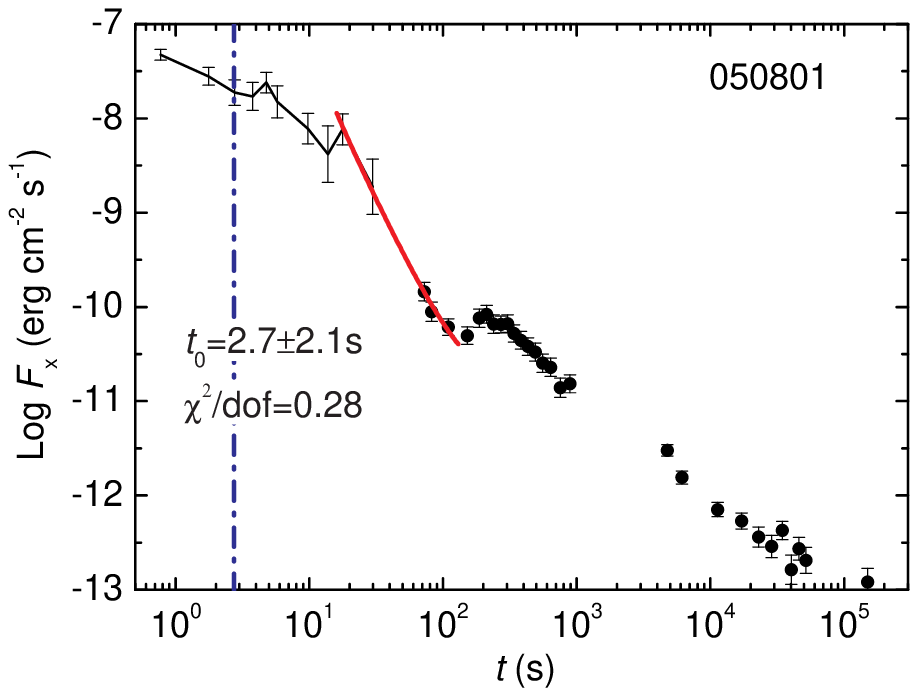}
\includegraphics[angle=0,scale=.50]{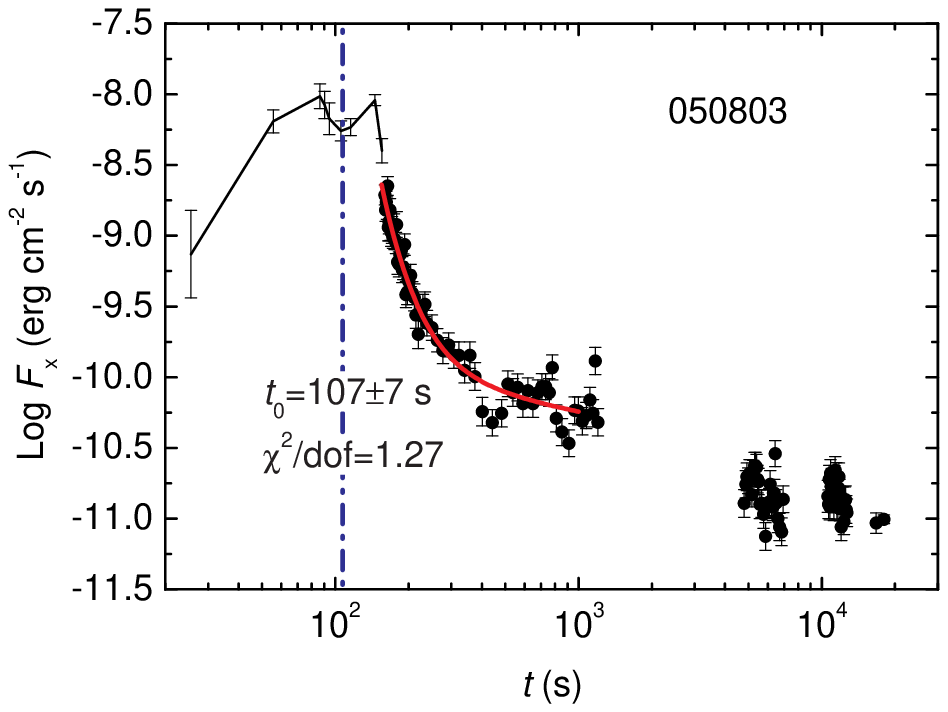}
\includegraphics[angle=0,scale=.50]{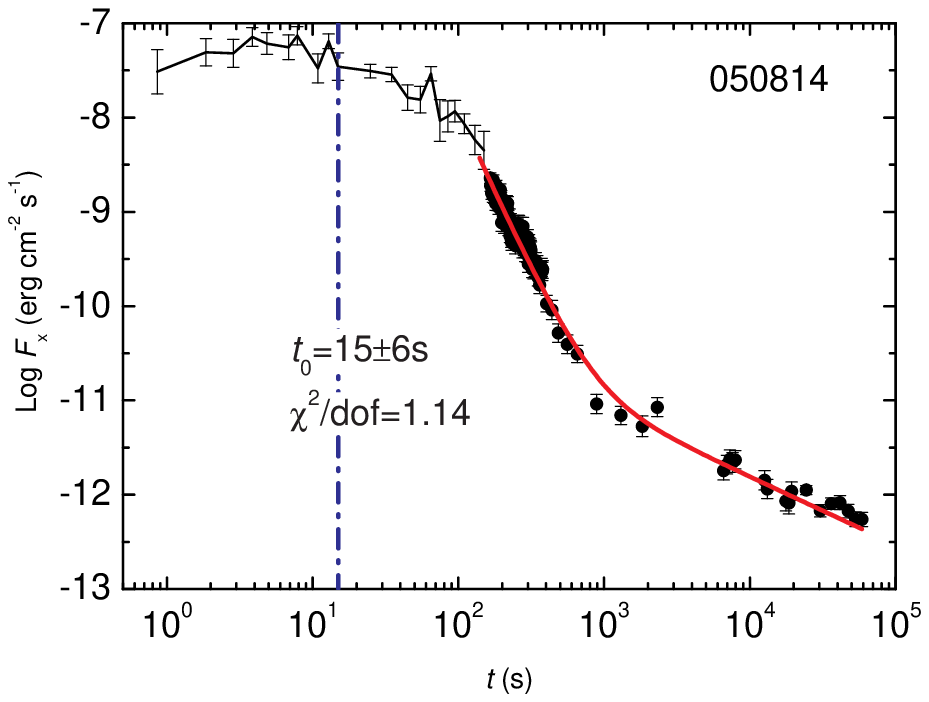}
\includegraphics[angle=0,scale=.50]{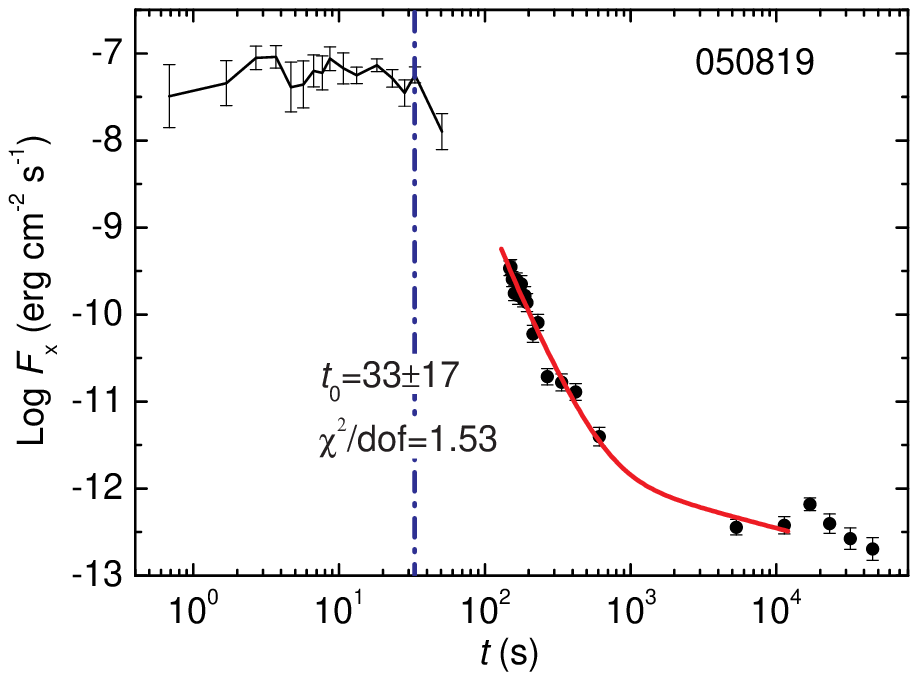}
\includegraphics[angle=0,scale=.50]{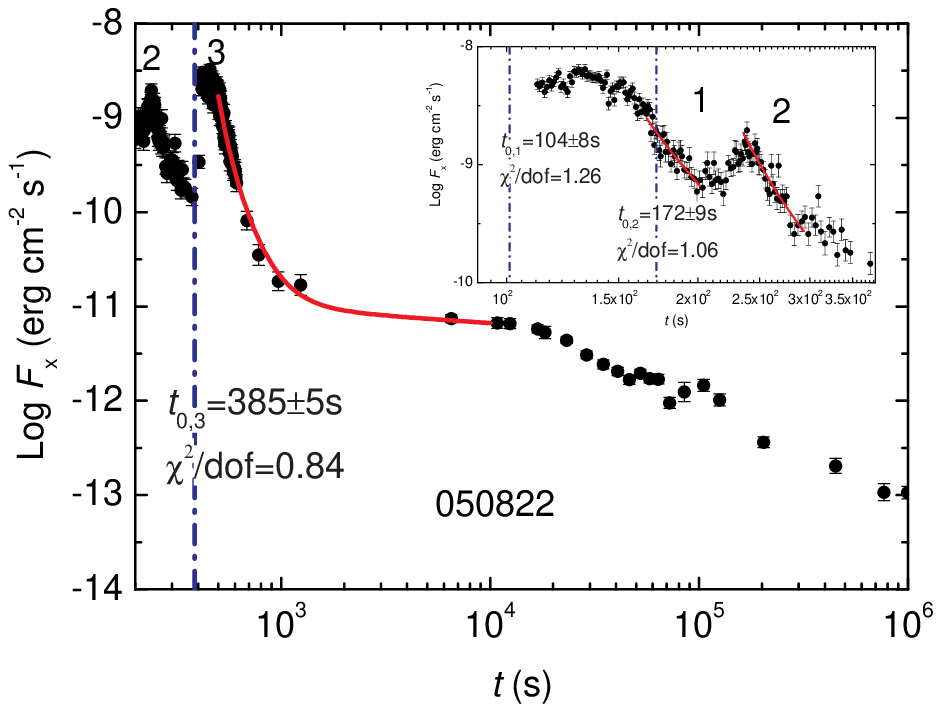}

{Fig 1. continued}
\end{figure}

\begin{figure}
\plotone{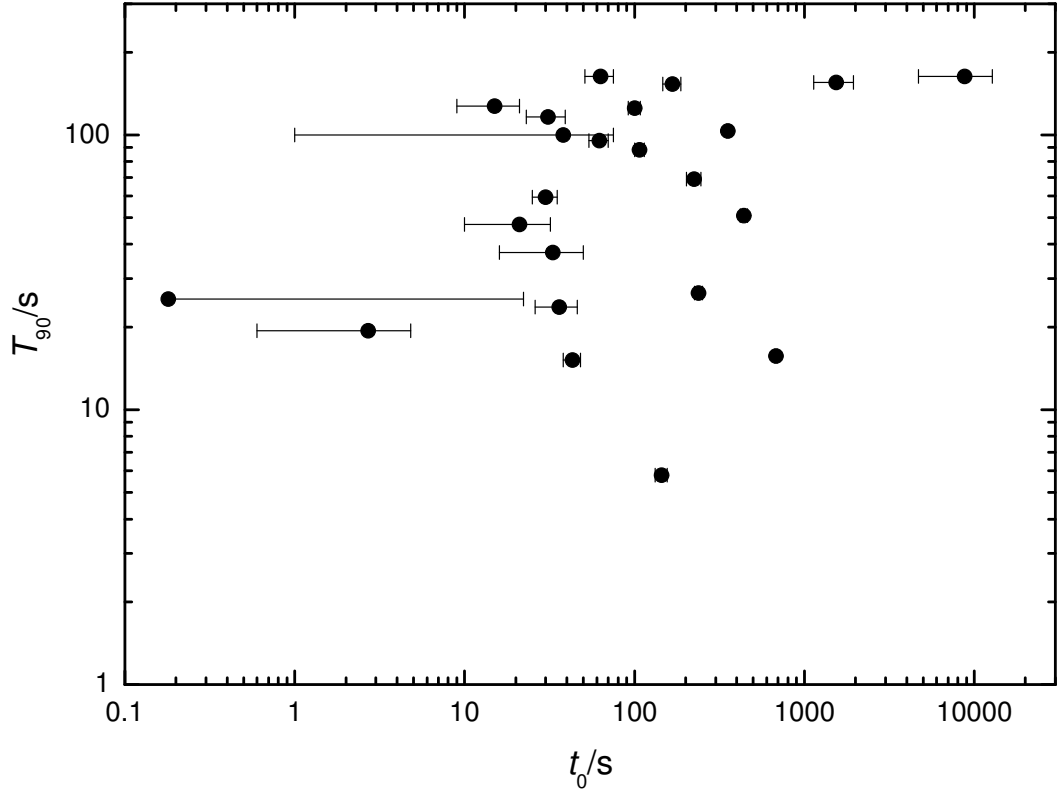} \caption{The $t_0$ as a function of $T_{90}$ for the tails/flares in our
sample, except for those heavily overlapped flares in the early time XRT light curves.
\label{fig2}}
\end{figure}

\end{document}